\documentclass[journal]{IEEEtran}

\usepackage{orcidlink}
\usepackage{cite}

\usepackage{amsmath,amssymb,amsfonts}
\usepackage{bm}

\usepackage{graphicx}
\usepackage{textcomp}
\usepackage{xcolor}
\usepackage{tikz}

\usepackage{algorithm}
\usepackage{algpseudocode}

\usepackage{tabularx}
\usepackage{booktabs}
\usepackage{multirow}
\usepackage{pifont}

\usepackage{comment}

\usepackage[caption=false,font=footnotesize]{subfig}

\usepackage{dblfloatfix}

\usepackage{placeins}

\hypersetup{hidelinks}

\begin{document}
\bstctlcite{IEEEcontrol}

\title{AngularWM: Wireless World Modeling for AoA Prediction from Multi-Antenna I/Q}

\author{ Omar~Mashaal,~\IEEEmembership{Student~Member,~IEEE}, 
        and Hatem~Abou-Zeid,~\IEEEmembership{Member,~IEEE}%

\thanks{ This work was supported in part by Alberta Innovates and in part by the Natural Sciences and Engineering Research Council of Canada (NSERC) under Grant 2021-04050.
(Corresponding author: Hatem~Abou-Zeid.)}%
\thanks{The authors are with the Department of Electrical and Software Engineering,
University of Calgary, Calgary, AB T2N 1N4, Canada
(e-mail: omar.mashaal1, hatem.abouzeid@ucalgary.ca).}
}

\maketitle

\begin{abstract}

Tracking angle-of-arrival (AoA) under mobility enables proactive beam alignment in multi-antenna wireless systems. However, learning this evolution directly from raw I/Q data is challenging because multipath, channel-gain variations, hardware impairments, and noise mix with angular information. This paper proposes \emph{AngularWM}, a self-supervised wireless world-modeling approach that learns a predictive latent angular state from ordered multi-antenna I/Q observations. A joint-embedding predictive architecture is combined with a label-free trajectory-level geometric regularizer that encourages an approximately affine, non-collapsed latent evolution motivated by the antenna-array response. No AoA, motion, or displacement labels are used during pretraining. The resulting structured world state supports direct multi-step encoder-space extrapolation and generalizes to angular step sizes not seen during training. On large-scale over-the-air measurements, AngularWM reduces the mean predictor-free AoA error over eight steps from $29.97^\circ$ to $0.49^\circ$. After pretraining, only three or four labeled reference angles are required to calibrate the complete 225-point AoA grid, achieving $0.48^\circ$ and $0.43^\circ$ MAE, respectively. AngularWM also maintains robust beam alignment and spectral efficiency under increasing angular displacement, while reducing the deployed model size from 1.42M to 0.30M parameters and forecast latency.

\end{abstract}


\section{Introduction}

Multi-antenna wireless systems rely on spatial information to maintain reliable directional transmission and reception. Under user mobility, the angle of arrival (AoA) of a dominant propagation component evolves over time, and outdated angular information can lead to beam misalignment and degraded link quality \cite{aoa_tracking_1}. Tracking this evolution and forecasting future AoA can therefore support proactive beam management and other mobility-aware radio functions. Rather than estimating each angular state independently, a predictive model can exploit the temporal structure in recent received observations to anticipate how the spatial state will evolve \cite{aoa_tracking_2}.

Raw multi-antenna I/Q measurements provide an information-rich observation of this evolution. The relative amplitudes and phases
across antenna elements contain a direction-dependent spatial signature associated with AoA \cite{van2002optimum}. However, the received samples simultaneously include waveform-dependent variations, time-varying path gains, multipath contributions, hardware effects, and noise. Learning directly from these observations avoids the need for an intermediate channel representation or explicit AoA estimate as the model input. The challenge is that this learning requires the representation to separate the evolving angular state from other variations in the received signal.

Predictive representation learning provides a natural approach to this problem. Latent world models seek compact internal states that capture the evolution of dynamic environments \cite{ha2018world,lecun2022path}. In the wireless setting considered here, this corresponds to learning a compact latent radio state whose evolution reflects the underlying spatial dynamics of the received signal. Joint-embedding predictive architectures (JEPAs) learn representations by predicting target embeddings instead of reconstructing the complete observation \cite{assran2023ijepa}. Recent work has extended predictive latent learning to raw multi-antenna wireless observations and channel-state representations \cite{chu2026wirelessjepa,bouchaaya2025latent,yang2026jepa,gao2026jepacfm,luo2026csi}. For predictive AoA tracking, however, accurate latent prediction alone does not ensure that the encoder representation itself exhibits a simple and regular evolution. An expressive neural predictor may achieve low prediction error by modeling complex, nonlinear latent dynamics, without requiring the encoder to organize angular evolution into an approximately linear form.

To facilitate direct extrapolation from the learned encoder representation, we hypothesize that the structure of the antenna-array response can guide the encoder toward a more regular latent trajectory. In array spatial-frequency coordinates, which describe the phase progression across the antenna elements, a change in direction produces a structured change in these relative phases \cite{van2002optimum}. Specifically, successive spatial-frequency changes produce phase transformations that combine according to the accumulated displacement. This compositional structure motivates learning a latent representation in which angular evolution follows a simple and consistent progression. A training objective can therefore be introduced to encourage this geometric regularity using only the temporal ordering of received I/Q observations, without access to AoA labels or spatial-frequency displacements.

Motivated by this observation, we propose AngularWM, a self-supervised framework for learning a \emph{structured}  latent representation of angular evolution directly from ordered raw multi-antenna I/Q observations. AngularWM combines latent future prediction with a trajectory-level geometric regularizer that jointly fits the encoder states within an ordered observation window to an approximately affine path. A minimum-step constraint discourages trajectory collapse, while variance--covariance regularization promotes feature diversity across samples. The geometric objective uses only observation ordering and requires no AoA, motion-direction, spatial-frequency, or displacement labels during pretraining. The resulting encoder representation enables low-complexity extrapolation of future latent states without a learned predictor at inference, and calibration to physical AoA using only a few labeled reference
locations.

To the best of our knowledge, this is the first work to present AoA-centered wireless world modeling directly from ordered raw multi-antenna I/Q observations. A central contribution of the proposed framework is an antenna-manifold-motivated training objective that encourages angular evolution to follow approximately affine latent trajectories, enabling low-complexity AoA prediction with limited labeled calibration data. The main contributions are summarized as follows:

\begin{itemize}
\item  We propose a self-supervised framework for predictive AoA tracking directly from temporally ordered raw multi-antenna I/Q observations. The framework learns representations that capture angular evolution without requiring AoA, motion-direction, or displacement labels during pretraining, eliminating the need for intermediate channel estimates or explicit AoA estimates as model inputs.

\item We propose a label-free trajectory-level geometric regularizer that exploits the structured phase evolution of the antenna-array response to encourage approximately affine latent dynamics. The proposed objective jointly fits encoder states across each complete observation window, while a minimum-step constraint prevents trajectory collapse and variance--covariance regularization promotes feature diversity. This formulation explicitly shapes the encoder geometry to support latent-state extrapolation using only temporal ordering as supervision.
\item We show that the proposed framework enables low-complexity prediction of future angular states through direct extrapolation of encoder outputs, reducing inference complexity and eliminating the need for the JEPA predictor at deployment. AngularWM reduces the mean predictor-free AoA error over an eight-step horizon from $29.97^\circ$ to $0.49^\circ$. We further demonstrate label-efficient AoA decoding over the full $K=225$ angular grid using only three and four labeled anchor angles, linking the self-supervised representation to physical AoA with limited calibration.
\item We evaluate AngularWM against \emph{JEPA-P} and \emph{AngularWM-L} through latent-geometry diagnostics, multi-step AoA forecasting, few-anchor calibration, beamforming loss, and spectral efficiency. Training-objective ablations assess the contribution of each loss term, while additional evaluations examine generalization to larger angular displacements and quantify inference complexity and latency. \end{itemize}

The rest of this paper is organized as follows. Section~\ref{sec:related_work} reviews related work. Section~\ref{sec:system_model} presents the system model and predictive AoA tracking problem. Section~\ref{sec:proposed_method} introduces the proposed structured-latent AoA predictor. Sections~\ref{sec:experimental_setup} and~\ref{sec:results} present the experimental setup and results, respectively. Section~\ref{sec:cncl} concludes the paper.

\begin{figure*}[!t]
    \centering
    \includegraphics[width=0.9\textwidth]{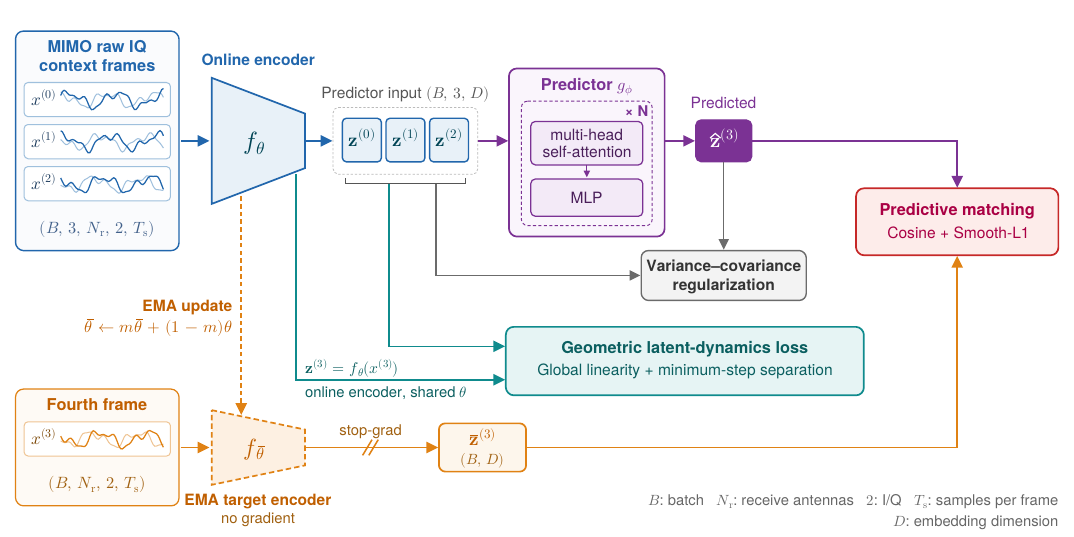}
    \caption{Overview of the proposed AngularWM framework with geometry-regularized JEPA learning.}
    \label{fig:jepa_framework}
\end{figure*}

\section{Related Work} 
\label{sec:related_work}

\subsection{Latent World Models and Predictive Learning for Wireless Signals}

Joint-embedding predictive architectures (JEPAs) learn representations by predicting target embeddings from context rather than reconstructing observations, allowing unpredictable observation-level details to be suppressed~\cite{lecun2022path}. I-JEPA demonstrated this principle for images~\cite{assran2023ijepa}, while V-JEPA extended it to video representation learning~\cite{bardes2024vjepa}. V-JEPA 2 further introduced an action-conditioned latent model for forecasting and planning~\cite{assran2025vjepa2}.

Collapse-prevention strategies include architectural asymmetry and stop-gradient updates, as well as explicit representation-level regularization such as VICReg and Barlow Twins~\cite{grill2020byol,bardes2022vicreg,zbontar2021barlow}.

Self-supervised representation learning has also gained increasing importance at the wireless physical layer. Masked and predictive objectives have been applied to spectrograms, channel state information (CSI), and raw wireless measurements for spectrum sensing, positioning, beam prediction, and other downstream tasks ~\cite{aboulfotouh2025wavesfm,aboulfotouh2025multimodal,yang2025wirelessgpt}. For raw I/Q data, IQFM showed that a common encoder can support multiple downstream radio tasks without intermediate handcrafted representations~\cite{mashaal2026iqfm}. More recently, JEPA-style predictive learning has been investigated for wireless signals and channel representations ~\cite{chu2026wirelessjepa,mohamed2026latentwave, gao2026jepacfm}. These studies highlight the value of latent prediction for transferable wireless representations, although explicit organization of the latent space according to the underlying physical state is generally not their primary objective.

\subsection{Structured Latent Geometry for Wireless Dynamics}

Structured latent geometry can support forecasting and planning. JEPA has been used to predict wireless propagation dynamics from CSI~\cite{bouchaaya2025latent}, while JEPA-MSAC applies temporal block-masked prediction to multimodal sensing and communication measurements~\cite{zheng2026jepamsac}. An action-conditioned CSI world model further imposes homomorphic, Lie-algebra-derived transitions to reflect user motion and spatial organization~\cite{naoumi2026homomorphic}. These approaches study predictive wireless dynamics using CSI or multimodal observations.

Channel charting provides a complementary approach to learning spatially organized representations from CSI~\cite{studer2018channel}. In visual world models, temporal straightening encourages consecutive latent velocities to align through a pairwise cosine criterion~\cite{wang2026temporal}. Our AngularWM-L baseline adapts this local regularization and additionally encourages consistent neighboring step magnitudes. AngularWM instead jointly fits a single affine path across the complete trajectory window and includes a minimum-step constraint to prevent trajectory collapse.

AngularWM combines predictive learning from ordered raw multi-antenna I/Q with complete-window trajectory regularization for AoA tracking. Without AoA or displacement labels during pretraining, it learns an angular representation that supports predictor-free extrapolation and few-anchor AoA calibration.

\section{System Model and Predictive AoA Tracking Problem}
\label{sec:system_model}

\subsection{Multi-Antenna Signal and Angular State}
\label{subsec:signal_model}

We consider a receiver equipped with a uniform planar array (UPA) of $N_r=N_hN_v$ antenna elements, where $N_h$ and $N_v$ denote the number of horizontal and vertical elements, respectively. The inter-element spacing is assumed to be half a wavelength. At time index $t$, the receiver records a raw multi-antenna I/Q frame
\begin{equation}
\mathbf X_t \in \mathbb{R}^{2\times N_r \times T_s},
\label{eq:raw_iq_frame}
\end{equation}
where the first dimension indexes the in-phase and quadrature components, and $T_s$ is the number of temporal samples per antenna. The observation $\mathbf{X}_t$ is processed directly from the received multi-antenna samples, without requiring an explicit channel estimate or a conventional AoA estimate as an intermediate input to the proposed model. For physical interpretation, we denote the corresponding complex-baseband I/Q frame observation by $\mathbf{Y}_t\in\mathbb{C}^{N_r\times T_s}$.

To characterize the AoA-dependent spatial structure, we consider the dominant-path geometric model

\begin{equation}
    \mathbf{Y}_t
    =
    \beta_t \mathbf{a}(\mathbf{q}_t)\mathbf{s}_t^{T}
    + \mathbf{U}_t ,
    \label{eq:dominant_path_model}
\end{equation}

where $\beta_t\in\mathbb{C}$ denotes the time-varying complex gain of the dominant propagation component, $\mathbf{s}_t\in\mathbb{C}^{T_s}$ is the transmitted waveform over the observation frame, and $\mathbf{a}(\mathbf{q}_t)\in\mathbb{C}^{N_r}$ is the array response associated with the dominant AoA. The aggregate term $\mathbf{U}_t$ represents received components that are not part of the dominant angular state, including receiver noise and, when present, non-dominant multipath, interference, and other unmodelled propagation effects. The physical state of interest in this work is therefore the direction of the dominant propagation component, while the raw I/Q observation may additionally vary due to channel gain, waveform content, multipath, interference, hardware effects, and noise.

We parameterize the physical AoA
\begin{equation}
\boldsymbol{\theta}_t
=
\begin{bmatrix}
\theta_{\mathrm{az},t} &
\theta_{\mathrm{el},t}
\end{bmatrix}^{\mathsf T}
\end{equation}
using the array-native spatial-frequency coordinate
\begin{equation}
\mathbf q_t
=
\begin{bmatrix}
q_{h,t} &
q_{v,t}
\end{bmatrix}^{\mathsf T}
=
\boldsymbol{\psi}
\left(
\boldsymbol{\theta}_t
\right),
\label{eq:q_coordinate}
\end{equation}
where $\boldsymbol{\psi}(\cdot)$ is an array-dependent mapping from physical azimuth and elevation to normalized spatial frequency. Its components, $q_{h,t}$ and $q_{v,t}$, describe the phase progression along the horizontal and vertical array axes, respectively. The exact form of $\boldsymbol{\psi}(\cdot)$ depends on the array orientation and angular coordinate convention.

For half-wavelength element spacing, the response of the element at
position $(m,n)$ is
\begin{equation}
\left[\mathbf a(\mathbf q_t)\right]_{\ell(m,n)}
=
\frac{1}{\sqrt{N_hN_v}}
\exp\left(
j\pi
\left[
m q_{h,t}
+
n q_{v,t}
\right]
\right),
\label{eq:upa_response}
\end{equation}
where $m=0,\ldots,N_h-1$, $n=0,\ldots,N_v-1$, and $\ell(m,n)$ maps the two-dimensional array position to its corresponding vector index. In general, the mapping $\boldsymbol{\psi}(\cdot)$ is nonlinear in the physical angles. Consequently, equal increments in azimuth or elevation do not necessarily correspond to equal spatial-frequency increments. Over a sufficiently short and
smooth angular trajectory, however, the array-native state varies smoothly, motivating a local approximation of its evolution rather than an assumption of global angular linearity.

The spatial-frequency state $\mathbf q_t$ is introduced only to describe the physical structure underlying the multi-antenna observations. Neither $\mathbf q_t$ nor the displacement between successive angular states is provided during self-supervised pretraining. Instead, the model is trained solely from ordered raw-I/Q trajectories
\begin{equation}
\mathcal T_b
=
\left\{
\mathbf X_b^{(0)},
\mathbf X_b^{(1)},
\mathbf X_b^{(2)},
\mathbf X_b^{(3)}
\right\},
\label{eq:ordered_trajectory}
\end{equation}
where the superscript specifies the observation order within the trajectory, and the four frames correspond to successive locations along an underlying angular evolution. For the experiments, these sequences are formed from the available I/Q measurements to represent successive angular states. The model receives only the resulting I/Q sequence during pretraining; AoA values, spatial-frequency coordinates, and inter-frame displacements are not provided.

\subsection{Array-Manifold Structure of Angular Motion}
\label{subsec:array_structure}
The array manifold provides the physical motivation for imposing a structured geometry on the learned latent space. For a UPA with half-wavelength spacing, a displacement in spatial frequency,
\begin{equation}
\Delta\mathbf q
=
\begin{bmatrix}
\Delta q_h & \Delta q_v
\end{bmatrix}^{\mathsf T},
\end{equation}
induces a deterministic phase-gradient change across the array. In particular, the steering vector at the displaced state can be written as
\begin{equation}
\mathbf a(\mathbf q+\Delta\mathbf q)
=
\mathbf D(\Delta\mathbf q)\mathbf a(\mathbf q),
\label{eq:array_action}
\end{equation}
where $\mathbf D(\Delta\mathbf q)$ is a diagonal phase-shift operator
with diagonal entry
\begin{equation}
\left[
\mathbf D(\Delta\mathbf q)
\right]_{\ell(m,n),\ell(m,n)}
=
\exp
\left(
j\pi
\left[
m\Delta q_h+n\Delta q_v
\right]
\right).
\label{eq:phase_operator}
\end{equation}
Hence, a change in AoA does not produce an arbitrary transformation of the multi-antenna observation; it produces a structured phase displacement determined by the array geometry.

Successive spatial-frequency displacements compose additively:
\begin{equation}
\mathbf D(\Delta\mathbf q_1)
\mathbf D(\Delta\mathbf q_2)
=
\mathbf D(\Delta\mathbf q_1+\Delta\mathbf q_2).
\label{eq:D_composition}
\end{equation}
This property shows that angular motion has a simple additive structure in the array-native coordinate, even though the raw I/Q observations vary nonlinearly with AoA through complex exponentials and are further affected by waveform, gain, multipath, and noise.

This motivates learning an encoder
\begin{equation}
\mathbf z_t = f_{\theta}(\mathbf X_t),
\end{equation}
whose latent space reflects the low-dimensional geometry of the underlying angular state. Since the raw observation also contains nuisance factors beyond AoA, we describe the desired angular organization through the mean embedding associated with each spatial-frequency state:
\begin{equation}
\boldsymbol{\mu}_z(\mathbf q)
=
\mathbb E
\left[
f_{\theta}(\mathbf X)
\mid
\mathbf q
\right].
\end{equation}
Nearby AoA states that form ordered curves or grids in the physical angular coordinate should therefore induce similarly ordered mean embeddings, rather than being mapped to arbitrary or scattered latent locations.

For small angular displacements, this desired behavior can be viewed through the first-order approximation
\begin{equation}
\boldsymbol{\mu}_z(\mathbf q+\Delta\mathbf q)
-
\boldsymbol{\mu}_z(\mathbf q)
\approx
\mathbf J_z(\mathbf q)\Delta\mathbf q,
\label{eq:local_latent_linearity}
\end{equation}

where $\mathbf J_z(\mathbf q)$ is the Jacobian of the mean embedding with respect to the spatial-frequency coordinate. In a generic representation, the latent response to the same angular motion may vary with location, producing curved or direction-dependent trajectories. Rather than requiring the embedding to preserve the generally nonlinear mapping from physical AoA to spatial frequency, we seek a simpler latent coordinate system in which ordered angular evolution is represented approximately affinely. For an ordered trajectory parameterized by $\tau$, the desired behavior is

\begin{equation}
\boldsymbol{\mu}_{z}\!\left(\mathbf q(\tau)\right)
\approx
\mathbf c + \tau \mathbf v,
\label{eq:global_latent_affine}
\end{equation}
where $\mathbf c$ and $\mathbf v$ denote the trajectory intercept and latent direction, respectively. Here, $\tau$ denotes the relative position within the ordered observation sequence. The affine relation therefore encourages successive angular observations to follow an approximately regular progression in the latent space, facilitating simple extrapolation. This formulation does not require $\mathbf q$ to vary linearly with the physical AoA. Instead, the array manifold provides the physical motivation that angular motion induces structured changes in the multi-antenna observation, while the learned representation is encouraged to linearize this evolution. Applying this constraint across trajectories spanning different locations and directions encourages a more consistent angular organization over the sampled AoA region. Since neither $\mathbf q$ nor its displacement is available during self-supervised pretraining, the corresponding physical structure cannot be imposed directly. We therefore impose the latent trajectory constraint using only the ordering of the raw-I/Q observations, as described in the following subsections.

\subsection{AoA World Model Formulation}
\label{subsec:problem_formulation}

Given a sequence of the most recent multi-antenna observations,
\begin{equation}
    \mathcal{X}_t
    =
    \{
    \mathbf X_{t-K+1},
    \ldots,
    \mathbf X_t
    \},
    \label{eq:context_set}
\end{equation}
the objective is to learn a representation of the angular state and model its evolution from observation history.
An encoder maps each frame to a latent state $\mathbf z_i=f_\theta(\mathbf X_i)$ intended to capture its
angular condition. The AoA world model uses the recent state history to forecast future states $\widehat{\mathbf z}_{t+h}$, which are decoded into physical azimuth and elevation estimates,
\begin{equation}
    \widehat{\boldsymbol{\theta}}_{t+h},
    \qquad h=1,\ldots,H.
    \label{eq:future_aoa_problem}
\end{equation}
In the evaluated implementation, $K=3$ recent I/Q frames form the observed context. During self-supervised pretraining, the corresponding AoA values, spatial-frequency coordinates, motion directions, and state displacements are not supplied to the model.

Future angular information is relevant to directional multi-antenna operation because an estimate $\widehat{\boldsymbol{\theta}}_{t+h}$ can be used to select or prepare a future steering or combining direction. Angular prediction errors can cause beam misalignment, reducing received signal quality and achievable spectral efficiency. The downstream communication impact is evaluated in Section~\ref{sec:results}.

We therefore seek a learned angular state with two complementary properties. First, it should be \emph{predictive}, retaining the information required to infer future angular states from recent I/Q observations. Second, its geometry should organize angular evolution into approximately affine latent trajectories that support \emph{constant-velocity extrapolation}. The proposed framework in Section~\ref{sec:proposed_method} is designed to jointly promote these properties.

\section{Proposed Structured-Latent AoA Predictor}
\label{sec:proposed_method}

\subsection{JEPA-Based Latent Predictive Learning}
\label{subsec:jepa_learning}

As illustrated in Fig.~\ref{fig:jepa_framework}, we instantiate a joint embedding predictive architecture to learn predictive latent states from ordered raw-I/Q trajectories. For each trajectory $\mathcal T_b$, the first three frames form the context, while the fourth frame is treated as the next-state prediction target. The model therefore predicts the latent representation of the next observation along the ordered angular trajectory rather than reconstructing its raw waveform.

The online encoder maps the context frames to latent states
\begin{equation}
\mathbf z_b^{(k)}
=
f_{\theta}
\left(
\mathbf X_b^{(k)}
\right),
\qquad
k=0,1,2.
\label{eq:context_embeddings}
\end{equation}
A target encoder $f_{\bar{\theta}}$ maps the target frame to
\begin{equation}
\bar{\mathbf z}_b^{(3)}
=
\operatorname{sg}
\left[
f_{\bar{\theta}}
\left(
\mathbf X_b^{(3)}
\right)
\right],
\label{eq:target_embedding}
\end{equation}
where $\operatorname{sg}[\cdot]$ denotes stop-gradient. Gradients are not propagated through the target branch. Instead, the target-encoder parameters are updated as an exponential moving average of the online encoder:
\begin{equation}
\bar{\theta}
\leftarrow
m\bar{\theta}
+
(1-m)\theta,
\label{eq:ema_update}
\end{equation}
where $m\in[0,1)$ is the momentum coefficient.

A predictor $g_{\phi}$ receives the ordered context embeddings and estimates the target representation:
\begin{equation}
\hat{\mathbf z}_b^{(3)}
=
g_{\phi}
\left(
[
\mathbf z_b^{(0)},
\mathbf z_b^{(1)},
\mathbf z_b^{(2)}
]
\right).
\label{eq:predictor}
\end{equation}
The prediction objective combines cosine dissimilarity and
Smooth-$L_1$ matching:
\begin{equation}
\mathcal L_{\rm JEPA}
=
\lambda_{\rm cos}\mathcal L_{\rm cos}
+
\lambda_{\rm s1}\mathcal L_{\rm s1},
\label{eq:jepa_loss}
\end{equation}
where
\begin{equation}
\mathcal L_{\rm cos}
=
\frac{1}{B}
\sum_{b=1}^{B}
\left(
1-
\frac{
\hat{\mathbf z}_b^{(3)\mathsf T}
\bar{\mathbf z}_b^{(3)}
}{
\left\|\hat{\mathbf z}_b^{(3)}\right\|_2
\left\|\bar{\mathbf z}_b^{(3)}\right\|_2
+\epsilon
}
\right),
\label{eq:cosine_prediction_loss}
\end{equation}
and
\begin{equation}
\mathcal L_{\rm s1}
=
\frac{1}{B}
\sum_{b=1}^{B}
\operatorname{SmoothL1}
\left(
\hat{\mathbf z}_b^{(3)},
\bar{\mathbf z}_b^{(3)}
\right).
\label{eq:smooth_l1_prediction_loss}
\end{equation}

The predictive objective encourages the online encoder to retain information useful for estimating the next latent state. For AoA-driven trajectories, this information is expected to depend strongly on the spatial phase relationships across the antenna array, which evolve systematically with angular motion. Prediction alone, however, does not guarantee a globally consistent angular geometry. An expressive predictor may compensate for curved or location-dependent latent trajectories while still accurately estimating the target embedding. We therefore complement predictive learning with a label-free trajectory-geometry regularizer.

\subsection{Label-Free Trajectory-Level Geometry Regularization}
\label{subsec:trajectory_geometry}

To evaluate the geometry of the complete trajectory, all four frames, including the prediction target, are encoded using the online encoder:
\begin{equation}
\mathbf z_b^{(k)}
=
f_{\theta}
\left(
\mathbf X_b^{(k)}
\right),
\qquad
k=0,1,2,3.
\label{eq:all_online_embeddings}
\end{equation}
Thus, the fourth frame is encoded by the EMA target branch for predictive matching and by the online branch for geometric regularization. The affine-path constraint can alternatively be applied to the three online context embeddings together with the stop-gradient EMA target embedding.

The objective is to encourage all four embeddings within each ordered trajectory to follow a single affine path. Unlike a local constraint that compares only adjacent increments, the proposed formulation fits the complete trajectory window jointly.

For trajectory $b$, define the centered order indices
\begin{equation}
\tau_k
=
k-\frac{3}{2},
\qquad k=0,1,2,3,
\end{equation}
such that
\begin{equation}
\tau_k\in
\left\{
-\frac{3}{2},
-\frac{1}{2},
\frac{1}{2},
\frac{3}{2}
\right\},
\qquad
\sum_{k=0}^{3}\tau_k=0,
\qquad
\sum_{k=0}^{3}\tau_k^2=5.
\end{equation}
We fit an affine path to the four embeddings:
\begin{equation}
\mathbf z_{{\rm fit},b}^{(k)}
=
\mathbf c_b^{\star}
+
\tau_k\mathbf v_b^{\star},
\label{eq:fitted_latent_line}
\end{equation}
where the fitted intercept and trajectory slope are
\begin{equation}
\mathbf c_b^{\star}
=
\frac{1}{4}
\sum_{k=0}^{3}
\mathbf z_b^{(k)},
\label{eq:centered_intercept}
\end{equation}
and
\begin{equation}
\mathbf v_b^{\star}
=
\frac{
\sum_{k=0}^{3}
\tau_k\mathbf z_b^{(k)}
}{
\sum_{k=0}^{3}\tau_k^2
}.
\label{eq:centered_velocity}
\end{equation}
The line-fitting loss penalizes deviations from the fitted path:
\begin{equation}
\mathcal L_{\rm lin}
=
\frac{1}{4B}
\sum_{b=1}^{B}\sum_{k=0}^{3}
\left\|
\mathbf z_b^{(k)}-\mathbf z_{{\rm fit},b}^{(k)}
\right\|_2.
\label{eq:global_linear_loss}
\end{equation}
Because the order indices are equally spaced, this affine parameterization encourages both trajectory straightness and approximately equal latent increments.

This constraint differs from pairwise temporal smoothness, which compares only neighboring increments, such as $\mathbf z_b^{(1)}-\mathbf z_b^{(0)}$ and $\mathbf z_b^{(2)}-\mathbf z_b^{(1)}$. Such a local constraint may improve short-range continuity without requiring all four embeddings to be explained by one fitted affine line. In contrast, \eqref{eq:global_linear_loss} penalizes deviations over the complete ordered window, which is important for reliable latent extrapolation.

Linearity alone does not prevent a collapsed trajectory, because four identical embeddings also produce zero line-fitting residual. We therefore define the consecutive latent increments
\begin{equation}
\mathbf d_b^{(k)}
=
\mathbf z_b^{(k+1)}
-
\mathbf z_b^{(k)},
\qquad
k=0,1,2,
\label{eq:latent_increment}
\end{equation}
and their lengths
\begin{equation}
s_b^{(k)}
=
\left\|
\mathbf d_b^{(k)}
\right\|_2.
\end{equation}
The minimum-step separation loss is
\begin{equation}
\mathcal L_{\rm step}
=
\frac{1}{3B}
\sum_{b=1}^{B}
\sum_{k=0}^{2}
\left[
\gamma_s-s_b^{(k)}
\right]_+,
\label{eq:min_step_loss}
\end{equation}
where $\gamma_s>0$ is a latent step-size margin and $[x]_+=\max(x,0)$. This term does not encode a physical angular distance or force all trajectories to have the same scale. It only prevents successive embeddings from collapsing to the same latent location. The minimum-step term is applied to ordered trajectories with nonzero angular progression and serves only as an anti-collapse constraint; it does not specify the physical magnitude of the angular displacement.

The complete trajectory-geometry regularizer is

\begin{equation}
\mathcal L_{\rm geom}
=
\lambda_{\rm lin}\mathcal L_{\rm lin}
+
\lambda_{\rm step}\mathcal L_{\rm step}.
\label{eq:geom_loss}
\end{equation}

The line-fitting term encourages each complete trajectory to follow an affine path, while the minimum-step term ensures that the path has nonzero extent. Both terms use only the ordering of raw-I/Q frames and require no AoA labels.

\subsection{Variance--Covariance Regularization and Complete Objective}
\label{subsec:vicrec_objective}

Predictive matching and line fitting alone do not ensure a diverse representation. The minimum-step loss penalizes small consecutive latent displacements, but does not directly control featurewise variance or correlations across the batch. We therefore use the variance and covariance components of VICReg to encourage feature diversity and reduce linear redundancy.

Both penalties are evaluated separately on the predicted target embeddings and the pooled online-encoder embeddings of the three context frames. Define
\begin{equation}
\begin{aligned}
\widehat{\mathbf Z}
&=
\left[
\widehat{\mathbf z}_1^{(3)},\ldots,
\widehat{\mathbf z}_B^{(3)}
\right]^{\mathsf T},\\
\mathbf Z_{\mathrm{ctx}}
&=
\left[
\mathbf z_1^{(0)},\mathbf z_1^{(1)},\mathbf z_1^{(2)},
\ldots,
\mathbf z_B^{(0)},\mathbf z_B^{(1)},\mathbf z_B^{(2)}
\right]^{\mathsf T},
\end{aligned}
\end{equation}
where $\widehat{\mathbf Z}\in\mathbb R^{B\times D}$ and $\mathbf Z_{\mathrm{ctx}}\in\mathbb R^{3B\times D}$. The fourth online embedding is used for geometric regularization, while the EMA target embedding provides the stop-gradient target for predictive matching; neither is included in these matrices.

For an embedding matrix $\mathbf Z\in\mathbb R^{N\times D}$ with $N>1$, let $\overline Z_j=N^{-1}\sum_{n=1}^{N}Z_{nj}$ denote the mean of column $j$. Using sample variance, the stabilized standard deviation and variance penalty are
\begin{equation}
\begin{aligned}
s_j(\mathbf Z)
&=
\sqrt{
\frac{1}{N-1}
\sum_{n=1}^{N}(Z_{nj}-\overline Z_j)^2
+\epsilon_{\rm var}
},\\
\mathcal L_{\rm var}(\mathbf Z)
&=
\frac{1}{D}
\sum_{j=1}^{D}
\left[1-s_j(\mathbf Z)\right]_+,
\end{aligned}
\label{eq:var_loss}
\end{equation}
where $\epsilon_{\rm var}=10^{-4}$ and $[x]_+=\max(x,0)$. This penalty encourages variation in each latent dimension across the rows of $\mathbf Z$.

Let $\tilde{\mathbf Z}$ denote the column-centered matrix. The sample covariance matrix is
\begin{equation}
\operatorname{Cov}(\mathbf Z)
=
\frac{1}{N-1}
\tilde{\mathbf Z}^{\mathsf T}\tilde{\mathbf Z},
\end{equation}
and the covariance penalty is
\begin{equation}
\mathcal L_{\rm cov}(\mathbf Z)
=
\frac{1}{D}
\sum_{i\neq j}
\left[
\operatorname{Cov}(\mathbf Z)
\right]_{ij}^{2}.
\label{eq:cov_loss}
\end{equation}
This penalty discourages linear redundancy by penalizing off-diagonal covariance entries.

The two penalties are computed within each embedding matrix, without concatenating the prediction and context batches. The resulting regularizer is
\begin{equation}
\begin{aligned}
\mathcal L_{\rm VC}
={}&
\lambda_{\rm var}
\left[
\mathcal L_{\rm var}(\widehat{\mathbf Z})
+
\mathcal L_{\rm var}(\mathbf Z_{\mathrm{ctx}})
\right]\\
&+
\lambda_{\rm cov}
\left[
\mathcal L_{\rm cov}(\widehat{\mathbf Z})
+
\mathcal L_{\rm cov}(\mathbf Z_{\mathrm{ctx}})
\right],
\end{aligned}
\label{eq:vicrec_loss}
\end{equation}

The predictive baseline and complete AngularWM objectives are
\begin{equation}
\begin{aligned}
\mathcal L_{\rm base}
&= \mathcal L_{\rm JEPA}+\mathcal L_{\rm VC},\\
\mathcal L_{\rm AngularWM}
&= \mathcal L_{\rm base}+\mathcal L_{\rm geom},
\end{aligned}
\label{eq:total_loss}
\end{equation}

where $\lambda_{\rm lin}=\lambda_{\rm step}=0.05$. The predictive term encourages future-state information, the geometric terms favor approximately affine trajectories with nonzero successive displacements, and the variance--covariance regularizer encourages featurewise diversity with reduced linear redundancy.

\begin{algorithm}[t]
\caption{Self-Supervised Training of AngularWM}
\label{alg:jepa_g}
\begin{algorithmic}[1]
\Require Four-frame I/Q sequences
$\{(\mathbf x_b^{(0)},\ldots,\mathbf x_b^{(3)})\}_{b=1}^{B}$;
online encoder $f_\theta$;
EMA target encoder $f_{\bar\theta}$;
predictor $g_\phi$
\Require $\lambda_{\rm var}=0.05$,
$\lambda_{\rm cov}=0.005$,
$\lambda_{\rm lin}=0.05$,
$\lambda_{\rm step}=0.05$,
$\gamma_s=0.3$;
scheduled EMA momentum $m_t\in[0.998,0.9999]$

\State $\bar\theta\gets\theta$
\For{each training minibatch at iteration $t$}
    \For{$k=0,1,2$}
        \State $\mathbf z_b^{(k)}
        \gets f_\theta(\mathbf x_b^{(k)})$,
        $b=1,\ldots,B$
    \EndFor

    \State $\widehat{\mathbf z}_b^{(3)}
    \gets g_\phi(
    \mathbf z_b^{(0)},\mathbf z_b^{(1)},\mathbf z_b^{(2)})$

    \State $\bar{\mathbf z}_b^{(3)}
    \gets\operatorname{sg}
    [f_{\bar\theta}(\mathbf x_b^{(3)})]$

    \State $\mathcal L_{\rm JEPA}
    \gets\mathcal L_{\rm cos}+0.05\,\mathcal L_{\rm s1}$
    \hfill $(\beta=1)$

    \State $\widehat{\mathbf Z}
    \gets\operatorname{stack}_{b}
    \{(\widehat{\mathbf z}_b^{(3)})^{\mathsf T}\}$

    \State $\mathbf Z_{\rm ctx}
    \gets\operatorname{stack}_{b,k}
    \{(\mathbf z_b^{(k)})^{\mathsf T}\}$,
    $k=0,1,2$

    \State Compute $\mathcal L_{\rm VC}$
    using \eqref{eq:vicrec_loss}

    \State $\mathbf z_b^{(3)}
    \gets f_\theta(\mathbf x_b^{(3)})$

    \State $\mathbf Z_{{\rm geo},b}\gets
    [\mathbf z_b^{(0)},\mathbf z_b^{(1)},
     \mathbf z_b^{(2)},\mathbf z_b^{(3)}]^{\mathsf T}$

    \State Fit least-squares affine paths to
    $\mathbf Z_{{\rm geo},b}$

    \State Compute $\mathcal L_{\rm lin}$
    (mean unsquared residual)

    \State Compute $\mathcal L_{\rm step}$
    (margin $\gamma_s$)

    \State $\displaystyle
    \begin{aligned}
    \mathcal L_{\rm AngularWM}\gets{}&
    \mathcal L_{\rm JEPA}+\mathcal L_{\rm VC}\\
    &+\lambda_{\rm lin}\mathcal L_{\rm lin}
    +\lambda_{\rm step}\mathcal L_{\rm step}
    \end{aligned}$

    \State Clear gradients; backpropagate
    $\mathcal L_{\rm AngularWM}$
    \State Clip the joint gradient norm
    \State Optimizer step for $\theta$ and $\phi$
    \State $\bar\theta\gets
    m_t\bar\theta+(1-m_t)\theta$
\EndFor
\end{algorithmic}
\end{algorithm}
\subsection{Local Trajectory Regularization}
\label{subsubsec:local_trajectory}

The AngularWM-L variant augments the baseline JEPA--VICReg objective with a local trajectory regularizer. For the four online-encoder embeddings, define the consecutive latent increments as
\begin{equation}
\mathbf d_b^{(k)}
=
\mathbf z_b^{(k+1)}
-
\mathbf z_b^{(k)},
\qquad
k=0,1,2.
\label{eq:local_increment}
\end{equation}

The local direction-consistency loss is
\begin{equation}
\mathcal L_{\mathrm{dir}}
=
\frac{1}{2B}
\sum_{b=1}^{B}
\sum_{k=0}^{1}
\left(
1-
\frac{
\left(\mathbf d_b^{(k)}\right)^{\mathsf T}
\mathbf d_b^{(k+1)}
}{
\|\mathbf d_b^{(k)}\|_2
\|\mathbf d_b^{(k+1)}\|_2
+\epsilon
}
\right),
\label{eq:local_direction_loss}
\end{equation}
which encourages neighboring increments to follow similar directions.

To encourage locally consistent step magnitudes, we also use
\begin{equation}
\mathcal L_{\mathrm{norm}}
=
\frac{1}{2B}
\sum_{b=1}^{B}
\sum_{k=0}^{1}
\rho_{0.1}
\left(
\|\mathbf d_b^{(k+1)}\|_2
-
\|\mathbf d_b^{(k)}\|_2
\right),
\label{eq:local_norm_loss}
\end{equation}
where $\rho_{0.1}(\cdot)$ denotes the Smooth-L1 penalty with
$\beta=0.1$.

The resulting local regularizer is
\begin{equation}
\mathcal L_{\mathrm{local}}
=
\mathcal L_{\mathrm{dir}}
+
\mathcal L_{\mathrm{norm}},
\label{eq:local_trajectory_loss}
\end{equation}
and the AngularWM-L objective is
\begin{equation}
\mathcal L_{\mathrm{AngularWM-L}}
=
\mathcal L_{\mathrm{base}}
+
\lambda_{\mathrm{local}}
\mathcal L_{\mathrm{local}},
\qquad
\lambda_{\mathrm{local}}=0.05.
\label{eq:jepa_local_objective}
\end{equation}

This regularizer is local because it compares only neighboring latent increments. In contrast, AngularWM fits a single affine path jointly to the complete four-frame trajectory.

\subsection{Proposed Few-Anchor AoA Decoding}
\label{subsec:few_anchor_decoding}

\begin{figure}[!t]
    \centering
    \includegraphics[width=0.95\columnwidth]
    {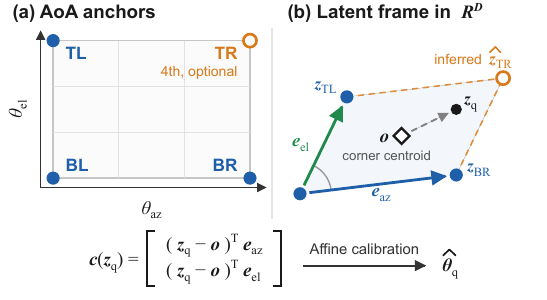}
    \caption{Few-anchor AoA decoding from three or four labeled corner
    embeddings. In the three-anchor case, the missing corner is inferred
    from the learned affine geometry before calibration.}
    \label{fig:anchor_axis_calibration}
\end{figure}

After self-supervised pretraining, the encoder is frozen. To exploit the angular structure induced by the proposed geometric regularization, we introduce a few-anchor decoder that maps learned embeddings to physical azimuth and elevation using only three or four labeled AoA locations. As illustrated in Fig.~\ref{fig:anchor_axis_calibration}, the decoder does not use PCA; instead, it defines the angular coordinate frame directly in the original $D$-dimensional embedding space.

Let
\begin{equation}
\mathbf z_{\mathrm{BL}},\;
\mathbf z_{\mathrm{BR}},\;
\mathbf z_{\mathrm{TL}},\;
\mathbf z_{\mathrm{TR}}
\in\mathbb R^D
\end{equation}
denote the embeddings associated with the bottom-left, bottom-right,
top-left, and top-right AoA-grid corners, respectively.

For the four-anchor decoder, all four corner embeddings are available.
We compute the boundary midpoints
\begin{equation}
\begin{aligned}
\mathbf z_{\mathrm L}
&=
\frac{\mathbf z_{\mathrm{BL}}+\mathbf z_{\mathrm{TL}}}{2},
&
\mathbf z_{\mathrm R}
&=
\frac{\mathbf z_{\mathrm{BR}}+\mathbf z_{\mathrm{TR}}}{2},
\\
\mathbf z_{\mathrm B}
&=
\frac{\mathbf z_{\mathrm{BL}}+\mathbf z_{\mathrm{BR}}}{2},
&
\mathbf z_{\mathrm T}
&=
\frac{\mathbf z_{\mathrm{TL}}+\mathbf z_{\mathrm{TR}}}{2}.
\end{aligned}
\label{eq:anchor_midpoints}
\end{equation}

The corresponding latent azimuth and elevation directions are
\begin{equation}
\mathbf e_{\mathrm{az}}
=
\frac{\mathbf z_{\mathrm R}-\mathbf z_{\mathrm L}}
{\|\mathbf z_{\mathrm R}-\mathbf z_{\mathrm L}\|_2},
\qquad
\mathbf e_{\mathrm{el}}
=
\frac{\mathbf z_{\mathrm T}-\mathbf z_{\mathrm B}}
{\|\mathbf z_{\mathrm T}-\mathbf z_{\mathrm B}\|_2}.
\label{eq:anchor_axes}
\end{equation}

These directions are normalized independently and need not be orthogonal. Any coupling caused by their nonorthogonality is handled by the subsequent affine calibration. The coordinate origin is the centroid of the four corner embeddings,
\begin{equation}
\mathbf o
=
\frac{
\mathbf z_{\mathrm{BL}}
+\mathbf z_{\mathrm{BR}}
+\mathbf z_{\mathrm{TL}}
+\mathbf z_{\mathrm{TR}}
}{4}.
\label{eq:anchor_origin}
\end{equation}

For any query embedding $\mathbf z$, its two-dimensional anchor-defined coordinate is
\begin{equation}
\mathbf c(\mathbf z)
=
\begin{bmatrix}
(\mathbf z-\mathbf o)^{\mathsf T}\mathbf e_{\mathrm{az}}
\\[2pt]
(\mathbf z-\mathbf o)^{\mathsf T}\mathbf e_{\mathrm{el}}
\end{bmatrix}.
\label{eq:anchor_coordinate}
\end{equation}

Using the labeled corner anchors
\begin{equation}
\mathcal A_4
=
\{\mathrm{BL},\mathrm{BR},\mathrm{TL},\mathrm{TR}\},
\label{eq:four_anchor_set}
\end{equation}
we fit an affine mapping from the anchor-defined latent coordinates to
physical AoA:
\begin{equation}
\widehat{\mathbf B}
=
\arg\min_{\mathbf B}
\sum_{a\in\mathcal A_4}
\left\|
\begin{bmatrix}
\mathbf c(\mathbf z_a)^{\mathsf T} & 1
\end{bmatrix}
\mathbf B
-
\boldsymbol{\theta}_a^{\mathsf T}
\right\|_2^2,
\label{eq:anchor_affine_fit}
\end{equation}
where
\begin{equation}
\boldsymbol{\theta}_a
=
\begin{bmatrix}
\theta_{\mathrm{az},a}\\
\theta_{\mathrm{el},a}
\end{bmatrix}
\in\mathbb R^2,
\qquad
\mathbf B\in\mathbb R^{3\times2}.
\end{equation}

The continuous AoA estimate for a query embedding $\mathbf z_q$ is then
\begin{equation}
\widehat{\boldsymbol{\theta}}_q
=
\left(
\begin{bmatrix}
\mathbf c(\mathbf z_q)^{\mathsf T} & 1
\end{bmatrix}
\widehat{\mathbf B}
\right)^{\mathsf T}.
\label{eq:decoded_angle}
\end{equation}

For the three-anchor setting, only three of the four corner embeddings are measured. Three non-collinear corners provide two independent latent directions and are therefore sufficient to define a two-dimensional angular coordinate system. The missing corner is inferred using the parallelogram relation implied by an affine latent geometry.

For example, when the top-right corner is omitted,
\begin{equation}
\widehat{\mathbf z}_{\mathrm{TR}}
=
\mathbf z_{\mathrm{BR}}
+
\mathbf z_{\mathrm{TL}}
-
\mathbf z_{\mathrm{BL}}.
\label{eq:three_anchor_missing_corner}
\end{equation}
Its corresponding physical coordinate follows the same affine relation,
\begin{equation}
\widehat{\boldsymbol{\theta}}_{\mathrm{TR}}
=
\boldsymbol{\theta}_{\mathrm{BR}}
+
\boldsymbol{\theta}_{\mathrm{TL}}
-
\boldsymbol{\theta}_{\mathrm{BL}}.
\label{eq:three_anchor_missing_angle}
\end{equation}

The inferred corner is then combined with the three measured anchors, and the same four-corner axis construction and affine decoder in \eqref{eq:anchor_midpoints}--\eqref{eq:decoded_angle} are applied. Importantly, the inferred corner does not constitute an additional labeled observation because both its latent representation and physical coordinate are determined entirely from the three available anchors.

Calibration is evaluated over 20 trials. In each trial, one measured embedding is selected from each of the four fixed grid corners. The four-anchor calibration method uses all four embeddings. The three-anchor calibration method is evaluated by omitting each corner in turn and averaging the resulting four mean absolute errors (MAEs).

\begin{figure}[!t]
    \centering
    \includegraphics[width=0.75\columnwidth]
    {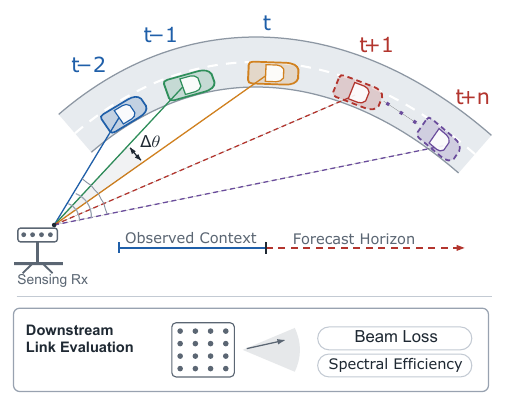}
    \caption{Forecasting and downstream link-level evaluation protocol
    using JEPA prediction or constant-velocity extrapolation.}
    \label{fig:forecast_link_protocol}
\end{figure}

\section{Experimental Setup and Evaluation Protocol}
\label{sec:experimental_setup}

\subsection{Over-the-Air Dataset and Data Split}
\label{subsec:measurement_dataset}

The experiments use over-the-air multi-antenna I/Q measurements collected with an outdoor software-defined-radio (SDR) testbed. The setup consists of one USRP X300 transmitter and two synchronized USRP X300 receivers, providing four receive channels in total. Synchronization is provided by an Ettus OctoClock CDA-2990. The radios operate at 5.88~GHz using microstrip planar antennas, and the two receiver units are separated by approximately 2.5~m. Measurements were collected over three outdoor sessions under naturally occurring multipath conditions.

The measurement corpus includes 16-QAM, 64-QAM, BPSK, QPSK, and PAM4 signals, together with sine and continuous-wave signals, captured at 1 and 10~MSps. Each
multi-channel I/Q frame has size $4\times2\times256$,
corresponding to four receive channels, I/Q components, and
256 temporal samples, respectively.

The measured angular region spans azimuth and elevation from $-70^\circ$ to $70^\circ$ in $10^\circ$ increments, yielding a $15\times15$ grid of 225 angular states. The configurations were generated through a mechanically controlled azimuth--elevation sweep during data collection.

The resulting dataset contains approximately five million four-frame I/Q sequences, covering diverse angular trajectories. Of these, $3{,}415{,}120$ sequences are used for predictive pretraining, while the remaining sequences form the held-out test set and are excluded from model training. Training and evaluation trajectories are constructed from
disjoint sets of I/Q frames, with no underlying frame shared
across partitions.

\subsection{Training Configuration and Compared Methods}

All models share a four-stage convolutional encoder producing $D=256$ embeddings and a two-layer, two-head Transformer predictor with learned positional embeddings and a query token. The EMA target encoder copies the online architecture and receives no gradient updates. The encoder and predictor contain $297{,}376$ and $1{,}121{,}792$ parameters, respectively, giving $1{,}419{,}168$ trainable parameters and $1{,}716{,}544$ including the EMA encoder. Training uses AdamW with batch size 512 and learning rates of $0.003$ for JEPA-P and $0.01$ for AngularWM-L and AngularWM. Gradient norms are clipped to $2.0$. The evaluated models share the same architecture. All models use the same held-out evaluation data and downstream protocols.

\subsection{Forecasting and Link-Level Evaluation}
\label{subsec:forecast_link_eval}

We compare recursive JEPA prediction with predictor-free constant-velocity extrapolation in the learned latent space. For the latter, the latent displacement is estimated from the two most recent observed online-encoder states:
\begin{equation}
\widehat{\mathbf v}_t
=
\mathbf z_t-\mathbf z_{t-1}.
\label{eq:latent_velocity}
\end{equation}
The $h$-step forecast is
\begin{equation}
\widehat{\mathbf z}_{t+h}
=
\mathbf z_t+h\widehat{\mathbf v}_t,
\qquad h=1,\ldots,H.
\label{eq:latent_extrapolation}
\end{equation}
Each horizon is obtained directly from the latest observed state using the same displacement estimate. In the angular-displacement experiment, the same rule is applied to strided observations with $h=1$. The overall forecasting and downstream evaluation protocol is illustrated in Fig.~\ref{fig:forecast_link_protocol}.

Forecast embeddings are converted to continuous AoA estimates using cosine-weighted $5$-nearest-neighbor regression. Encoder-only forecasts are decoded using a regressor fitted in the frozen online-encoder space, whereas Transformer-predicted embeddings are decoded using a separately fitted regressor in the target-encoder space.

Forecasting accuracy is measured by AoA MAE, defined at
each horizon as
$\operatorname{MAE}_h=
(\operatorname{MAE}_{{\rm az},h}
+\operatorname{MAE}_{{\rm el},h})/2$.
Figure~\ref{fig:encoder_extrapolation} reports the average
over horizons $1$--$8$, whereas
Table~\ref{tab:objective_ablation} reports horizon $8$.

To assess the downstream communication impact of AoA errors, we evaluate the predicted angles using a single-path $4\times4$ URA link model. The resulting beam-alignment and spectral-efficiency metrics quantify the link-level effect of the estimated AoA errors.

For azimuth $\phi$ and elevation $\theta$, define
\begin{equation}
u(\phi,\theta)=\sin(\phi)\cos(\theta),
\qquad
v(\theta)=\sin(\theta).
\end{equation}

For an $M_r\times M_c$ URA with half-wavelength spacing
($d_x=d_y=0.5$), the steering response at row $r$ and
column $c$ is
\begin{equation}
[\mathbf a(\phi,\theta)]_{r,c}
=
\exp\!\left(
j2\pi\left[
c d_x u(\phi,\theta)+r d_y v(\theta)
\right]
\right),
\label{eq:link_steering_vector}
\end{equation}

where $N=M_rM_c$.

The estimated AoA directly defines a continuous steering beam; no finite beam codebook is used for the reported beam-loss and spectral-efficiency results. The normalized beam-alignment gain is
\begin{equation}
g
=
\frac{
\left|
\mathbf a(\boldsymbol{\theta})^{\mathsf H}
\mathbf a(\widehat{\boldsymbol{\theta}})
\right|^2
}{
N^2
},
\qquad
0\leq g\leq1,
\label{eq:beam_alignment_gain}
\end{equation}
and the corresponding loss relative to perfect-angle steering is
\begin{equation}
L_{\mathrm{beam}}
=
-10\log_{10}g.
\label{eq:beam_loss}
\end{equation}

The effective SINR is modeled as
\begin{equation}
\gamma_{\mathrm{eff}}
=
\frac{N\gamma_0g\rho}{1+\eta},
\label{eq:effective_sinr}
\end{equation}
where $\gamma_0=10^{\mathrm{SNR}_0/10}$ is the per-element reference SNR, $\eta=10^{\mathrm{INR}/10}$ is the interference-to-noise ratio, and $\rho$ accounts for channel aging. We use $\mathrm{SNR}_0=0$~dB and $\mathrm{INR}=-10$~dB. Channel aging follows
\begin{equation}
\rho(\tau)
=
J_0^2(2\pi f_D\tau),
\label{eq:jakes_aging}
\end{equation}
where $J_0(\cdot)$ denotes the zeroth-order Bessel function of the first kind. The same aging model is applied to all methods. The resulting
spectral efficiency is
\begin{equation}
R
=
\log_2\left(1+\gamma_{\mathrm{eff}}\right)
\quad \mathrm{bit/s/Hz}.
\label{eq:spectral_efficiency}
\end{equation}

\subsection{Latent-Geometry Evaluation Metrics}
\label{subsec:geometry_metrics}

We evaluate the frozen online-encoder representation using five complementary geometry metrics reported in Table~\ref{tab:geometry}.

\paragraph{Trajectory straightness}
For trajectory $b$, let $\bar{\mathbf z}_b=\frac{1}{4}\sum_{k=0}^{3}\mathbf z_b^{(k)}$ and let $\widetilde{\mathbf Z}_b$ contain the rows $(\mathbf z_b^{(k)}-\bar{\mathbf z}_b)^{\mathsf T}$. Denoting the eigenvalues of $\widetilde{\mathbf Z}_b\widetilde{\mathbf Z}_b^{\mathsf T}$ by $\lambda_{b,1}\geq\cdots\geq\lambda_{b,4}$, we define
\begin{equation}
S_b=
\frac{\lambda_{b,1}}
{\sum_{j=1}^{4}\lambda_{b,j}+\epsilon}.
\label{eq:straightness_metric}
\end{equation}
Values near one indicate nearly linear trajectories. We average the score over at most $10{,}000$ trajectories.

\paragraph{Step cosine}
With $\mathbf d_b^{(k)}=\mathbf z_b^{(k+1)}-\mathbf z_b^{(k)}$ for $k=0,1$, consecutive-step alignment is measured by
\begin{equation}
C_b=
\frac{(\mathbf d_b^{(0)})^{\mathsf T}\mathbf d_b^{(1)}}
{\|\mathbf d_b^{(0)}\|_2\|\mathbf d_b^{(1)}\|_2+\epsilon}.
\label{eq:step_cosine_metric}
\end{equation}
Values near one indicate consistent latent directions.

\paragraph{Metric stress}
Let $\boldsymbol{\mu}_i$ denote the mean embedding at AoA grid point $i$, with distances $d_{ij}^{z}=\|\boldsymbol{\mu}_i-\boldsymbol{\mu}_j\|_2$ and $d_{ij}^{\theta}=\|\boldsymbol{\theta}_i-\boldsymbol{\theta}_j\|_2$. Using the fitted scale $\alpha^\star=\frac{\sum_{(i,j)}d_{ij}^{\theta}d_{ij}^{z}}{\sum_{(i,j)}(d_{ij}^{\theta})^2+\epsilon}$, normalized stress is
\begin{equation}
\mathrm{Stress}=
\sqrt{
\frac{\sum_{(i,j)}(d_{ij}^{z}-\alpha^\star d_{ij}^{\theta})^2}
{\sum_{(i,j)}(\alpha^\star d_{ij}^{\theta})^2+\epsilon}
}.
\label{eq:metric_stress}
\end{equation}
Lower values indicate better preservation of AoA distances. We use $50{,}000$ sampled grid-point pairs.

\paragraph{Jacobian condition number}
We estimate $\mathbf J_i$ using forward finite differences of the mean embedding along azimuth and elevation, divided by the corresponding grid spacings $h_{\mathrm{az}}$ and $h_{\mathrm{el}}$. For its largest and smallest singular values, $\sigma_{i,\max}$ and $\sigma_{i,\min}$, we define
\begin{equation}
\kappa_i=
\frac{\sigma_{i,\max}}
{\max(\sigma_{i,\min},\sqrt{\epsilon})}.
\label{eq:jacobian_condition}
\end{equation}
Values near one indicate low local anisotropy. We report the median over valid grid points.

\paragraph{Spearman distance correlation}
For sampled embedding pairs, let $d_{ij}^{\mathrm{cos}}=1-\frac{\mathbf z_i^{\mathsf T}\mathbf z_j}{\|\mathbf z_i\|_2\|\mathbf z_j\|_2}$. Agreement between the ordering of latent and physical AoA distances is measured by
\begin{equation}
\rho_s=
\operatorname{corr}_{\mathrm{rank}}
\left(d_{ij}^{\theta},d_{ij}^{\mathrm{cos}}\right).
\label{eq:spearman_metric}
\end{equation}
Values near one indicate strong agreement in distance ordering.

\begin{figure*}[!t]
    \centering
    \subfloat[JEPA-P]{
        \includegraphics[width=0.315\textwidth]
        {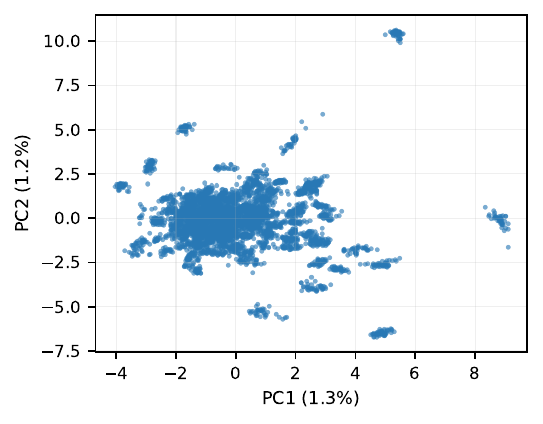}
    }
    \hfill
    \subfloat[AngularWM-L]{
        \includegraphics[width=0.315\textwidth]
        {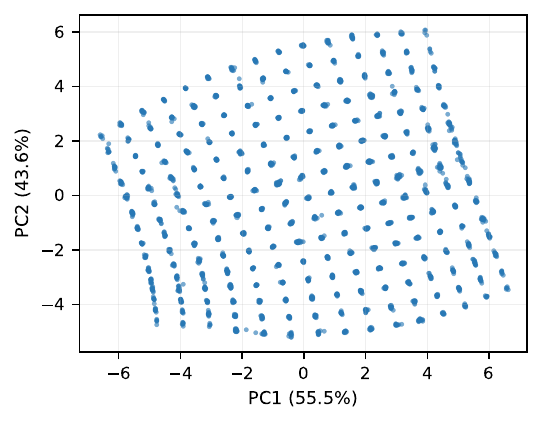}
    }
    \hfill
    \subfloat[AngularWM]{
        \includegraphics[width=0.315\textwidth]
        {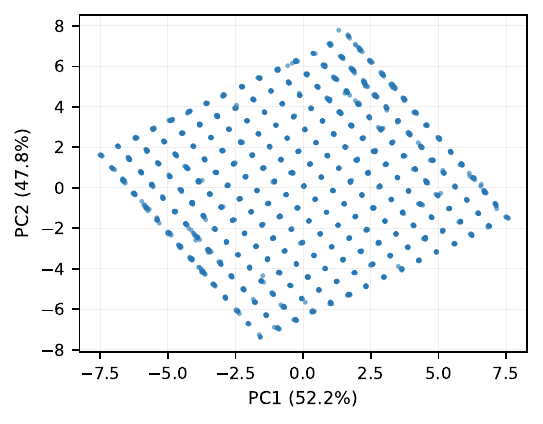}
    }

    \caption{PCA projections of the online-encoder embeddings over the AoA grid. JEPA-P shows no clear angular grid in the dominant PCA plane, whereas AngularWM-L and AngularWM exhibit ordered grid-like structures. Axis percentages indicate the variance explained by each principal component.}
    \label{fig:pca_geometry}
\end{figure*}

\section{Results and Discussion}
\label{sec:results}

We first examine how the different trajectory constraints organize the learned angular representation. We then evaluate whether the resulting geometry translates into accurate future-AoA forecasting, few-anchor physical calibration, and robust link-level performance. Finally, we analyze the contribution of the individual training objectives and the inference complexity of predictor-free forecasting.

\subsection{Geometry of the Learned Angular Representation}
\label{subsec:latent_geometry}

Figure~\ref{fig:pca_geometry} provides a qualitative view of the AoA-dependent organization of the online-encoder embeddings. PCA is used only to visualize the dominant geometric structure of each learned representation. For the baseline \emph{JEPA-P}, the projected embeddings form an irregular distribution with no clear correspondence to the angular grid in the dominant PCA plane. This indicates that predictive learning alone does not organize the encoder representation into a simple angular coordinate system.

The geometry-regularized variants exhibit a markedly different structure. \emph{AngularWM-L} produces an ordered latent arrangement, although the grid remains visibly sheared and similar angular displacements can correspond to different latent directions depending on their location in the AoA region. In contrast, \emph{AngularWM} yields a coherent grid-like organization in which neighboring AoA states remain ordered and the principal angular directions correspond to nearly consistent latent displacements. These observations suggest that the local constraint improves trajectory ordering, whereas the global trajectory-level constraint more effectively aligns angular evolution across different locations and directions.

\begin{figure}[!b]
    \centering
    \includegraphics[width=0.75\columnwidth]
    {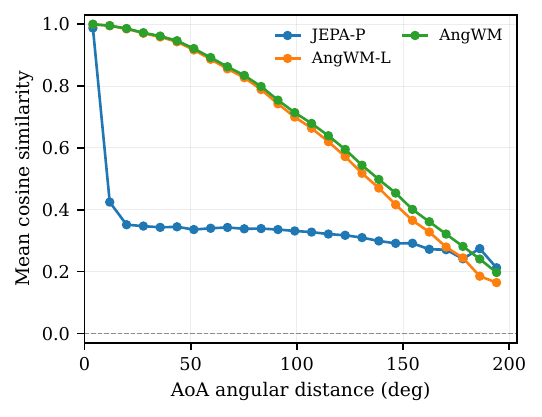}
    \caption{Mean cosine similarity between online-encoder embeddings as
    a function of the angular separation between their corresponding AoA
    states.}
    \label{fig:embedding_similarity_distance}
\end{figure}

Figure~\ref{fig:embedding_similarity_distance} examines how embedding similarity varies with physical AoA separation. For the baseline \emph{JEPA-P}, the mean cosine similarity drops sharply from nearly \(1\) at the smallest separations to about \(0.35\), then changes only gradually over the remaining angular range. This indicates that the baseline distinguishes nearly coincident AoA states but provides limited separation among moderately and widely separated locations. In contrast, the geometry-regularized variants exhibit a smooth, approximately monotonic decrease in similarity as angular separation increases, consistent with preservation of angular ordering. \emph{AngularWM-L} and \emph{AngularWM} show closely aligned trends, with \emph{AngularWM} achieving the strongest monotonic relationship. For compact presentation, \emph{AngularWM} is denoted as \emph{AngWM} in the figures. However, cosine similarity does not fully characterize Euclidean distance preservation, global affine consistency, or local directional anisotropy.

Table~\ref{tab:geometry} quantitatively compares the representations using the metrics defined in Section~\ref{subsec:geometry_metrics}. Both geometry-regularized variants substantially improve trajectory straightness and step-direction consistency over \emph{JEPA-P}. While local regularization already produces nearly straight trajectories with aligned latent increments, \emph{AngularWM} further improves angular-distance preservation and local isotropy, as reflected by its lower metric stress and Jacobian condition number. These results support the improved geometry obtained with the complete AngularWM objective.

\begin{table}[b]
\centering
\scriptsize
\setlength{\tabcolsep}{2.5pt}
\caption{Online-encoder geometry at $D=256$; arrows indicate the preferred direction.}
\label{tab:geometry}
\begin{tabular}{lccccc}
\toprule
Method & Straight. $\uparrow$ & Step cos. $\uparrow$ & Stress $\downarrow$ & Jac. cond. $\downarrow$ & Spearman $\rho$ $\uparrow$ \\
\midrule
JEPA-P & 0.424 & -0.391 & 0.449 & 1.577 & 0.233 \\
AngularWM-L & 0.999 & 0.994 & 0.065 & 1.129 & 0.990 \\
AngularWM & \textbf{1.000} & \textbf{0.998} & \textbf{0.006} & \textbf{1.022} & \textbf{0.996} \\
\bottomrule
\end{tabular}
\end{table}

\subsection{Future-AoA Forecasting from the Learned Representation}
\label{subsec:latent_extrapolation}

We compare recursive forecasting using the learned JEPA predictor with predictor-free constant-velocity extrapolation directly in the online-encoder space. This comparison tests whether accurate forecasting depends on the predictor or whether the learned encoder representation itself supports simple extrapolation. Both approaches follow the decoding protocol described in Section \ref{subsec:forecast_link_eval}.

Table~\ref{tab:objective_ablation} shows that accurate latent prediction does not necessarily imply an extrapolatable encoder space. The baseline \emph{JEPA-P} achieves a step-8 predictor MAE below \(0.001^{\circ}\), despite its weak latent geometry. In comparison, \emph{AngularWM-L} and \emph{AngularWM} obtain predictor MAEs of \(4.11^{\circ}\) and \(0.90^{\circ}\), respectively. This shows that an expressive predictor can accurately model future latent states even when the encoder itself does not provide a simple angular coordinate system.

\begin{figure}[!b]
    \centering
    \includegraphics[width=0.75\columnwidth]
    {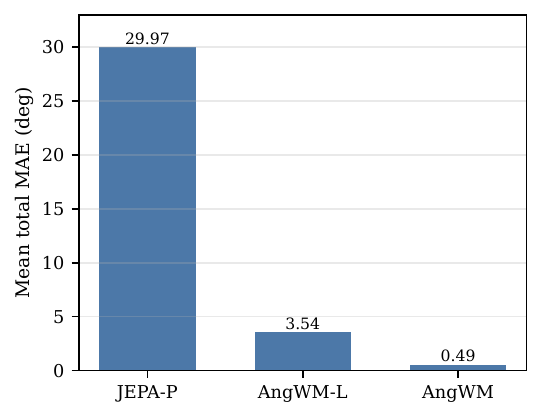}
    \caption{Mean AoA error for eight-step predictor-free linear
    extrapolation from the online-encoder embeddings.}
    \label{fig:encoder_extrapolation}
\end{figure}

The ranking changes substantially when the predictor is removed. Figure~\ref{fig:encoder_extrapolation} applies the same constant-velocity rule directly to the online-encoder embeddings of all methods and reports the AoA error averaged over the eight-step horizon. The baseline \emph{JEPA-P} produces a mean error of \(29.97^{\circ}\), showing that its accurate predictor rollout does not result from an encoder space that supports simple linear extrapolation. \emph{AngularWM-L} reduces the mean error to \(3.54^{\circ}\), while \emph{AngularWM} achieves \(0.49^{\circ}\), approximately \(61\times\) lower than the baseline under the same predictor-free extrapolation setting.

These results demonstrate that latent predictability and representation-level extrapolatability are distinct properties. The predictor can compensate for curved or location-dependent dynamics in the baseline latent space, whereas geometric regularization produces more consistent latent directions and increments that support forecasting without recursive predictor inference. In particular, \emph{AngularWM} provides highly accurate predictor-free extrapolation while retaining low predictor-based error.

\subsection{Few-Anchor AoA Calibration}
\label{subsec:anchor_decoding}

\begin{figure}[!tb]
    \centering
    \includegraphics[width=0.75\columnwidth]
    {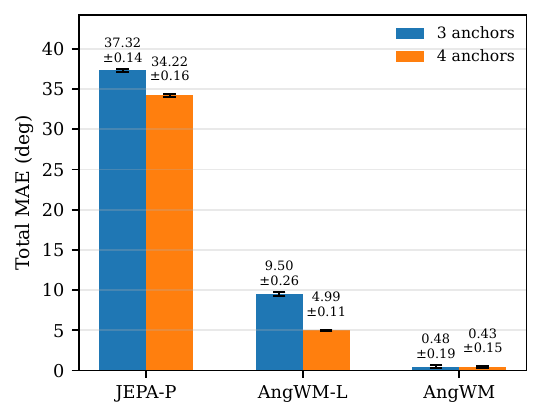}
\caption{AoA decoding error over the full 225-point angular grid using three or four labeled corner anchors.}
    \label{fig:anchor_decoding}
\end{figure}

We next evaluate whether the learned world-state representation is sufficiently organized to support decoding from only a few labeled AoA locations. Figure~\ref{fig:anchor_decoding} compares three- and four-anchor calibration over the complete angular grid. In the three-anchor setting, one corner is inferred from the remaining three according to the affine construction described in Section~\ref{subsec:few_anchor_decoding}.

The baseline \emph{JEPA-P} performs poorly under sparse calibration, with AoA MAEs of \(37.32^{\circ}\pm0.14^{\circ}\) and \(34.22^{\circ}\pm0.16^{\circ}\) using three and four anchors, respectively. Although its predictor accurately forecasts future embeddings, its irregular encoder geometry cannot be aligned with the physical angular coordinates using only a few affine reference points. \emph{AngularWM-L} reduces the corresponding errors to \(9.50^{\circ}\pm0.26^{\circ}\) and \(4.99^{\circ}\pm0.11^{\circ}\), showing that local trajectory regularization provides partial geometric organization.

The global trajectory-level constraint enables substantially more accurate full-grid calibration. \emph{AngularWM} achieves MAEs of \(0.48^{\circ}\pm0.19^{\circ}\) with three anchors and \(0.43^{\circ}\pm0.15^{\circ}\) with four anchors. The small degradation when one corner is inferred from the remaining three indicates that the learned latent geometry closely follows the intended affine organization. The anchors therefore do not define individual angular locations; rather, they calibrate a global angular geometry already learned from unlabeled raw-I/Q trajectories. These results demonstrate that the proposed representation supports highly label-efficient AoA decoding.

\subsection{Robustness to Angular Displacement and Link-Level Impact}
\label{subsec:mobility}

We next evaluate whether the learned angular geometry remains effective as the displacement between successive observations increases. For a fixed observation interval, this setting represents faster angular motion; equivalently, it represents more sparsely sampled observations along the same angular trajectory. For this link-level comparison, JEPA-P uses one predictor call, whereas AngularWM-L and AngularWM use
constant-velocity extrapolation.
\subsubsection{Robustness to Increasing Angular Displacement}

\begin{figure}[!t]
    \centering
    \includegraphics[width=0.75\columnwidth]
    {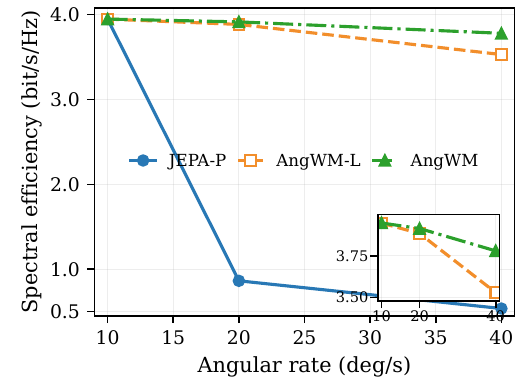}
    \caption{Spectral efficiency versus angular displacement
    per update.}
    \label{fig:angular_displacement_se}
\end{figure}

Figure~\ref{fig:angular_displacement_se} reports spectral efficiency as the angular displacement between successive observations increases. Trajectory strides of $s\in\{1,2,4\}$ correspond to displacements of approximately $10^\circ$, $20^\circ$, and $40^\circ$ (deg/step), respectively. At $10^\circ$, all three methods achieve approximately $3.947$ bit/s/Hz. At $40^\circ$, the spectral efficiency of \emph{JEPA-P} decreases to $0.537$ bit/s/Hz, whereas \emph{AngularWM-L} and \emph{AngularWM} retain $3.529$ and $3.779$ bit/s/Hz, respectively.

The results indicate that the geometry-regularized representations maintain effective extrapolation and beam alignment even when observations are widely separated along the angular trajectory, whereas the baseline predictor exhibits substantial degradation.

\subsubsection{Beam-Loss Reliability}

\begin{figure}[!t]
    \centering
    \includegraphics[width=0.8\columnwidth]
    {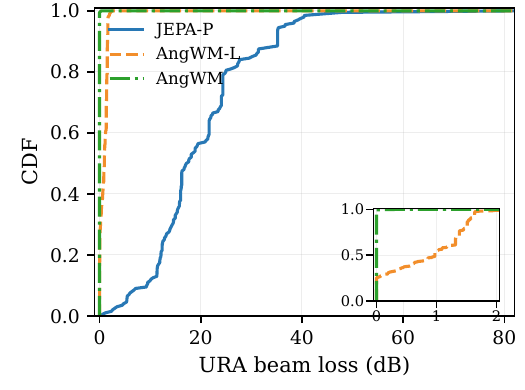}
    \caption{Empirical CDF of the synthesized $4\times4$ URA
    beamforming loss at the largest evaluated angular displacement
    of $40^\circ$, using the same $5{,}274$ evaluation samples
    for all methods.}
    \label{fig:beam_loss_ecdf}
\end{figure}

Figure~\ref{fig:beam_loss_ecdf} examines beam-alignment reliability at the largest evaluated angular displacement of $40^\circ$. The baseline \emph{JEPA-P} exhibits a broad beam-loss distribution, with mean, median, and 95th-percentile losses of $19.520$, $17.174$, and $36.659$ dB, respectively. In contrast, \emph{AngularWM-L} achieves mean, median, and 95th-percentile losses of $0.823$, $0.973$, and $1.636$ dB, respectively. No evaluated sample from either geometry-regularized method exceeds $3$ dB, and the maximum observed loss for \emph{AngularWM} is $0.339$ dB.

Geometric regularization enhances both average link performance and beam-alignment reliability by reducing the severe loss tail observed in the baseline. The tighter distribution achieved by \emph{AngularWM} further illustrates the advantages of globally consistent angular dynamics.

\begin{table*}[!bt]
\centering

\caption{Ablation of the prediction and geometry objectives.
MAEs are evaluated at step 8.}
\label{tab:objective_ablation}
\setlength{\tabcolsep}{3.2pt}
\renewcommand{\arraystretch}{1.12}
\begin{tabular}{lcccccc|ccccc}
\toprule
& \multicolumn{6}{c|}{Enabled objective terms}
& \multicolumn{1}{c}{Prediction}
& \multicolumn{2}{c}{Local geometry}
& \multicolumn{1}{c}{Global geometry}
& \multicolumn{1}{c}{Encoder-only} \\
\cmidrule(lr){2-7}
\cmidrule(lr){8-8}
\cmidrule(lr){9-10}
\cmidrule(lr){11-11}
\cmidrule(lr){12-12}
Method
& Cos.
& SL1
& VIC
& Local
& Global
& Step
& Pred. MAE
& Straight.
& Step Cos.
& Spearman
& Ext. MAE \\
&
&
&
&
&
&
&
($^\circ$) $\downarrow$
& $\uparrow$
& $\uparrow$
& $\uparrow$
& ($^\circ$) $\downarrow$ \\
\midrule
Cosine-only
& \checkmark & -- & -- & -- & -- & --
& 0.002 & 0.378 & -0.480 & 0.057 & 53.34 \\

Smooth-L1-only
& -- & \checkmark & -- & -- & -- & --
& 2.179 & 0.801 & 0.391 & 0.438 & 47.24 \\

JEPA w/o VICReg
& \checkmark & \checkmark & -- & -- & -- & --
& 2.756 & 0.792 & 0.367 & 0.476 & 47.44 \\

JEPA-P
& \checkmark & \checkmark & \checkmark & -- & -- & --
& $\boldsymbol{<0.001}$ & 0.424 & -0.391 & 0.233 & 53.33 \\

\midrule
AngularWM-L
& \checkmark & \checkmark & \checkmark & \checkmark & -- & --
& 4.111 & 0.999 & 0.994 & 0.990 & 8.82 \\

AngularWM w/o Step$^\dagger$
& \checkmark & \checkmark & \checkmark & -- & \checkmark & --
& 45.638 & 0.530 & -0.486 & -0.008 & 45.88 \\

\textbf{AngularWM}
& \checkmark & \checkmark & \checkmark & -- & \checkmark & \checkmark
& 0.896 & \textbf{1.000} & \textbf{0.998}
& \textbf{0.996} & \textbf{0.78} \\
\bottomrule
\end{tabular}

\vspace{1mm}
\footnotesize{
SL1: Smooth-L1; VIC: VICReg; Local/Global: trajectory regularization;
Step: minimum-step constraint.
$^\dagger$Collapsed representation; mean latent step norm
$2.04\times10^{-7}$.
}
\end{table*}

\subsection{Training-Objective Ablation}
\label{subsec:loss_ablation}

Table~\ref{tab:objective_ablation} shows that predictive accuracy and physically extrapolatable latent geometry are distinct properties. Cosine-only training achieves a step-8 predictor MAE of only \(0.002^\circ\), but produces weak geometry and a large \(53.34^\circ\) encoder-only extrapolation error. Smooth-L1 improves trajectory straightness, step consistency, and angular-distance correlation, but its extrapolation error remains high at \(47.24^\circ\). These results indicate that the prediction losses can produce highly forecastable embeddings without establishing a common angular coordinate system. VICReg complements these terms by preventing population-level degeneracy and supporting a stable predictive representation, rather than directly imposing angular geometry.

The complete predictive model, \emph{JEPA-P}, combines cosine, Smooth-L1, and VICReg and achieves a step-8 predictor MAE below \(0.001^\circ\). However, its negative step cosine of \(-0.391\), Spearman correlation of \(0.233\), and \(53.33^\circ\) encoder-only extrapolation error show that strong latent prediction alone does not produce a physically extrapolatable representation. \emph{JEPA-P} therefore provides a strong predictive reference and motivates the need for explicit geometric regularization. Adding local trajectory regularization in \emph{AngularWM-L} makes individual trajectories nearly linear, with straightness \(0.999\), step cosine \(0.994\), and Spearman correlation \(0.990\), reducing the extrapolation MAE to
\(8.82^\circ\).

The proposed \emph{AngularWM} further imposes consistency over the complete trajectory window, reaching straightness \(1.000\), step cosine
\(0.998\), and Spearman correlation \(0.996\), while reducing the step-8 encoder-only extrapolation MAE to \(0.78^\circ\). The minimum-step constraint is essential: removing it causes the globally fitted representation to collapse to a mean latent step norm of \(2.04\times10^{-7}\), with near-zero angular organization and an extrapolation MAE of \(45.88^\circ\). Overall, the ablation shows that the predictive objective provides a strong forecasting foundation, whereas the main contribution of \emph{AngularWM} is to transform this predictive representation into a non-collapsed, globally organized angular state that supports accurate predictor-free extrapolation.

\begin{figure}[!b]
    \centering
    \includegraphics[width=0.8\columnwidth]
        {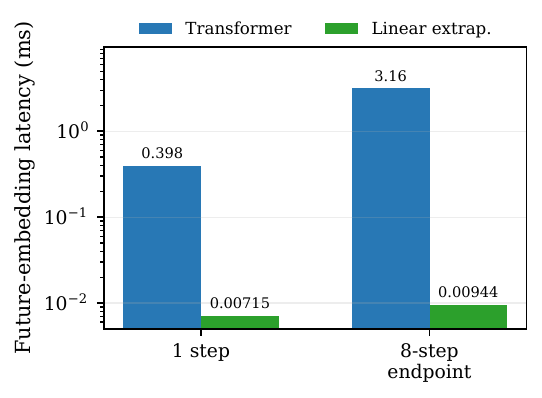}
    \caption{Forecast-only latency for recursive Transformer-predictor
    rollout and encoder-space linear extrapolation with cached context
    embeddings.}
    \label{fig:forecast_latency}
\end{figure}

\subsection{Inference Complexity and Latency}
\label{subsec:computational_efficiency}

The physical organization of the learned AoA representation enables a substantially simpler forecasting policy. Although all methods use the
JEPA predictor during training, the geometric regularizers introduce no additional inference-time operations and enable future embeddings to be extrapolated directly in the frozen online-encoder space. The learned Transformer predictor can therefore be replaced at deployment by simple vector arithmetic when the encoder geometry supports accurate linear extrapolation.

Figure~\ref{fig:forecast_latency} compares forecast-only latency when the required context embeddings are already cached. For one-step
forecasting, linear extrapolation requires \(0.0072\)~ms, compared with \(0.398\)~ms for the Transformer predictor, corresponding to
approximately \(55\times\) lower latency. At eight steps, recursive predictor rollout requires \(3.159\)~ms, whereas linear extrapolation
requires only \(0.0094\)~ms, yielding approximately \(336\times\) lower forecast latency.

Table~\ref{tab:controlled_complexity} summarizes the corresponding parameter and operation counts. A one-step Transformer-predictor evaluation requires \(6.46\) million profiled operations. In contrast, linear extrapolation requires \(D\) element-wise subtractions and \(D\) additions, totaling \(2D=512\) scalar operations for \(D=256\). Discarding the predictor also reduces the deployed model size from \(1.42\) million to \(0.30\) million parameters.

When one online-encoder evaluation is included, the one-step end-to-end latency decreases from \(1.042\) to \(0.651\)~ms. This reduction is smaller than the forecast-only gain because both policies incur the same encoder cost whenever a new I/Q frame is received. Multi-step forecasting, however, requires no additional encoder evaluations.

\begin{table}[t]
\centering
\caption{Inference complexity at $D=256$ (single-thread CPU, batch size 1).}
\label{tab:controlled_complexity}
\footnotesize
\setlength{\tabcolsep}{2.8pt}
\renewcommand{\arraystretch}{1.08}
\begin{tabular}{lccccc}
\toprule
Policy
& Params
& Ops$_1$
& $t_1$ (ms)
& $t_8$ (ms)
& $t_{\mathrm{e2e},1}$ (ms) \\
\midrule
Transformer pred.
& 1.42M & 6.46M & 0.398 & 3.159 & 1.042 \\

Linear extrap.
& 0.30M & 512 & 0.0072 & 0.0094 & 0.651 \\
\bottomrule
\end{tabular}
\end{table}
These results demonstrate that learning a physically extrapolatable angular representation provides both representational and computational benefits. The geometry-aware encoder supports accurate multi-step forecasting without recursive Transformer inference, substantially reducing forecast latency, deployed parameter count, and implementation complexity while retaining the same online encoder.

\section{Conclusion}
\label{sec:cncl}

This paper presented a structured-latent approach for predictive AoA tracking directly from raw multi-antenna I/Q observations. The proposed method combines JEPA-based latent prediction with a label-free geometric regularizer motivated by the antenna-array response, using only the ordering of the received observations during pretraining without requiring AoA labels or spatial-frequency displacement information. The results show that accurate latent prediction at the training angular step does not necessarily ensure robustness to changes in the inter-observation angular displacement, nor produce an encoder representation suitable for direct extrapolation. In contrast, the proposed geometric regularization produces a more organized angular representation that supports accurate multi-step predictor-free forecasting and full-grid AoA calibration from only three or four labeled anchors, while maintaining robust beam alignment as the angular displacement between observations increases. The learned geometry also enables removal of the predictor at deployment, reducing forecasting complexity. Overall, the results demonstrate the benefit of explicitly structuring the learned representation, in addition to optimizing predictive accuracy, for mobility-aware wireless prediction.

\bibliographystyle{IEEEtran}
\bibliography{name}

\end{document}